\documentclass[preprint]{aastex62}
\NewPageAfterKeywords

\shorttitle{}
\shortauthors{Gall et al.}

\begin{document}

\title{EBIT Observation of Ar Dielectronic Recombination Lines Near the Unknown Faint X-Ray Feature Found in the Stacked Spectrum of Galaxy Clusters}

\correspondingauthor{Amy C. Gall}
\email{acgall@g.clemson.edu}

\author{Amy C. Gall}
\affil{Clemson University, Department of Physics and Astronomy, Clemson, SC 29634-0978, USA}

\author{Adam R. Foster}
\affiliation{Harvard-Smithsonian Center for Astrophysics, 60 Garden Street, Cambridge, MA 02138, USA}

\author{Roshani Silwal}
\affiliation{Clemson University, Department of Physics and Astronomy, Clemson, SC 29634-0978, USA}
\affiliation{National Institute of Standards and Technology, Gaithersburg, MD 20899, USA}

\author{Joan M. Dreiling}
\affiliation{National Institute of Standards and Technology, Gaithersburg, MD 20899, USA}

\author{Alexander Borovik Jr.}
\affiliation{I. Physikalisches Institut, Justus-Liebig-Universität Gießen, 35392 Giessen, Germany}

\author{Ethan Kilgore}
\affiliation{Clemson University, Department of Physics and Astronomy, Clemson, SC 29634-0978, USA}

\author{Marco Ajello}
\affiliation{Clemson University, Department of Physics and Astronomy, Clemson, SC 29634-0978, USA}

\author{John D. Gillaspy}
\affiliation{National Institute of Standards and Technology, Gaithersburg, MD 20899, USA}
\affiliation{National Science Foundation, Alexandria, VA 22314, USA}

\author{Yuri Ralchenko}
\affiliation{National Institute of Standards and Technology, Gaithersburg, MD 20899, USA}

\author{Endre Tak\'acs}
\affiliation{Clemson University, Department of Physics and Astronomy, Clemson, SC 29634-0978, USA}
\affiliation{National Institute of Standards and Technology, Gaithersburg, MD 20899, USA}

\begin{abstract}
Motivated by possible atomic origins of the unidentified emission line detected at 3.55 keV to 3.57 keV in a stacked spectrum of galaxy clusters  \citep{Bulbul14}, an electron beam ion trap (EBIT) was used to investigate the resonant dielectronic recombination (DR) process in highly-charged argon ions as a possible contributor to the emission feature. The He-like Ar DR-induced transition 1s$^22$\emph{l} - 1s2\emph{l}3\emph{l}$^\prime$ was suggested to produce a 3.62 keV photon \citep{Bulbul14} near the unidentified line at 3.57 keV and was the starting point of our investigation. The collisional-radiative model NOMAD \citep{Ralchenko01} was used to create synthetic spectra for comparison with both our EBIT measurements and with spectra produced with the AtomDB database/Astrophysical Plasma Emission Code (APEC) \citep{Foster12,Smith01} used in the \cite{Bulbul14} work. Excellent agreement was found between the NOMAD and EBIT spectra, providing a high level of confidence in the atomic data used. Comparison of the NOMAD and APEC spectra revealed a number of missing features in the AtomDB database near the unidentified line. At an electron temperature of $T_e$ = 1.72 keV, the inclusion of the missing lines in AtomDB increases the total flux in the 3.5 keV to 3.66 keV energy band by a factor of 2. While important, this extra emission is not enough to explain the unidentified line found in the galaxy cluster spectra.       

\end{abstract}



\section{Introduction} \label{sec:intro}
Studies of galaxy clusters driven by the search for a dark matter candidate, the sterile neutrino, whose decay may produce an x-ray photon, have found a promising unidentified x-ray emission feature. The unknown feature has been reported at 3.55 keV to 3.57 keV \citep{Bulbul14} in the stacked X-ray Multi-Mirror (XMM-Newton) spectrum of high-count galaxy clusters and at 3.52 keV $\pm$ 0.02 keV \citep{Boyarsky14} in the XMM-Newton spectrum of the Perseus galaxy cluster and the Andromeda galaxy. \cite{Bulbul14} noted that the observed feature could be due to a number of atomic transitions including lines from Ar and K, while Gu et al. and Shah et al. made arguments in support of charge exchange between bare sulfur and atomic hydrogen occurring as a result of the interaction between the hot intracluster medium (ICM) and cold dense clouds in galaxy clusters \citep{Gu15,Shah16}. 

The possibility that the feature could be a signature of dark matter has spurred many follow-up studies: some confirmed the detection \citep{Urban15,Iakubovskyi15} while others, including the high-resolution broadband Hitomi results from the Perseus cluster \citep{Aharonian17}, found little evidence for the unidentified line \citep{Malyshev14,Anderson15,Tamura15,Carlson15,Sekiya16}. The existence of the unidentified line is still under investigation and may remain in question until future high-energy resolution x-ray satellite missions are able to measure the spectra with good energy resolution and sensitivity in a number of galaxy clusters. 

To help eliminate possible atomic origins and to aid the analysis of future observations near the unidentified line, we utilized the electron beam ion trap (EBIT) at the National Institute of Standards and Technology (NIST) to study the 1s$^2$2\emph{l} - 1s2\emph{l}3\emph{l}$^\prime$ resonant dielectronic recombination (DR) transitions in Li-like Ar (Ar$^{15+}$), which produce x-ray photons close in energy to the unknown line. In this work we show measured and calculated Ar x-ray spectra that include many DR satellites from lower charge-state ions that were not listed in AtomDB \citep{Foster12}, the atomic database that was used in the \cite{Bulbul14} analysis and often used in astrophysical x-ray spectral modeling. We further demonstrate that inclusion of these lines leads to a significant increase in emission in this energy region and produces agreement with measurements. 


\section{Dielectronic Recombination} \label{sec:DR}

Dielectronic recombination is a two-step resonant process, in which a free electron is captured into a bound state of an ion while an atomic electron is simultaneously propagated into an energetically higher bound state. The doubly excited ion then stabilizes through spontaneous decay, emitting a photon. The DR process is described by eqn. \ref{eq:1},

\begin{equation} \label{eq:1}
e^- + X^{q+} \rightarrow (X^{(q-1)+})^{**} \rightarrow X^{(q-1)+}+ h\nu
\end{equation}

where e$^-$ represents the free electron, X$^{q+}$ is an ion (X) with positive charge q$+$, (X$^{(q-1)+} )^{(**)}$ is the doubly-excited ion with charge (q-1)+, X$^{(q-1)+}$ is the stabilized ion, and $h\nu$ denotes an emitted photon. 

DR resonances are labeled using 3 letter notation, with the first, second, and third letter representing the principal quantum number of the initial unexcited bound electron, the excited electron, and the capture shell of the recombined electron, respectively. As a relevant example, during the KLM DR process, a free electron may be captured into the \emph{n} = 3 (M) shell while a bound electron is excited from \emph{n} = 1 (K) to \emph{n} = 2 (L). 

Dielectronic recombination can play an important role in determining the charge state balance of plasmas. This has motivated a number of EBIT and electron beam ion source (EBIS) measurements. For Ar$^{16+}$ in particular, measurements of cross-sections for DR on He-like Ar were performed by \cite{Ali90, Ali91}. These measurements were later expanded upon by \cite{Smith00} where good agreement was found between measurement and theory. Later EBIT measurements by \cite{Biedermann02} explored He-like Ar satellite lines for plasma temperature diagnostics.

\section{Experiment} \label{sec:exp}
X-ray spectra of highly-charged Ar ions were measured at the NIST EBIT facility. Its quasi-monoenergetic electron beam, with an energy spread and radius of approximately 50 eV and 35 $\mu$m, respectively, allows for ion charge-state and excitation selectivity \citep{Gillaspy96}. The electron beam is compressed to about a 10$^{11}$ cm$^{-3}$ density by a 2.7 T magnetic field produced by a pair of superconducting Helmholtz coils. The drift tube assembly, consisting of three sequentially aligned cylindrical tubes, traps ions axially while the space charge potential of the electron beam confines them radially. The voltages on the three drift tubes are floated on top of that of a shield electrode surrounding the drift tubes. The energy of the electrons in the interaction region is determined by the voltage of the middle drift tube, finely adjustable up to 30 kV, and the space charge of the electron beam \citep{Porto00}. Neutral atoms can be continuously injected into the interaction region using a ballistic gas injection system \citep{Fahy07} attached to one of the side ports oriented perpendicular to the electron beam. Additional ports located radially around the trap region are used for spectroscopic observations of the EBIT plasma. Presently, x-ray and EUV spectral regions can be accessed.  Further details of the design and operation of the NIST EBIT can be found in \citep{Gillaspy97}. 

For our investigation, neutral argon atoms were injected into the EBIT, and the electron beam current was set to 60 mA. The electron beam energy was initially set to 2.1 keV, well above the ionization threshold of Ar$^{15+}$ [918.375 eV from the NIST database \citep{Kramida18}]. The trap voltage cycle included a charge breeding time of 5 s followed by a 10 ms dumping interval to displace any buildup of contaminants such as barium ions sputtered out of the dispenser cathode of the electron gun. The measurements were performed in a steady-state mode where the electron beam energy was set to remain constant during measurements. In this mode, the EBIT plasma attains steady-state at each individual electron beam energy setting, and the charge-state balance at each energy can be properly accounted for by a non-Maxwellian collisional-radiative (CR) model. 

During our study, the electron beam energy was scanned from 2.1 keV to 5.2 keV in 15 eV steps to identify DR resonances. X-rays were collected for 3 minutes at each electron beam energy using a broad-band solid-state high purity germanium (HPGe) detector with 135 eV full width at half maximum (FWHM) energy resolution at 6.5 keV. Simultaneous measurements were taken with a high-resolution less than 2 eV FWHM at 3 keV) Johann-type crystal spectrometer \citep{Henins87} using a Si (111) crystal and an x-ray CCD detector. 

\section{ANALYSIS and RESULTS} \label{sec:results}
\subsection{Experimental Broadband Results} \label{subsec:broadband}
\begin{figure}[ht!]
\plotone{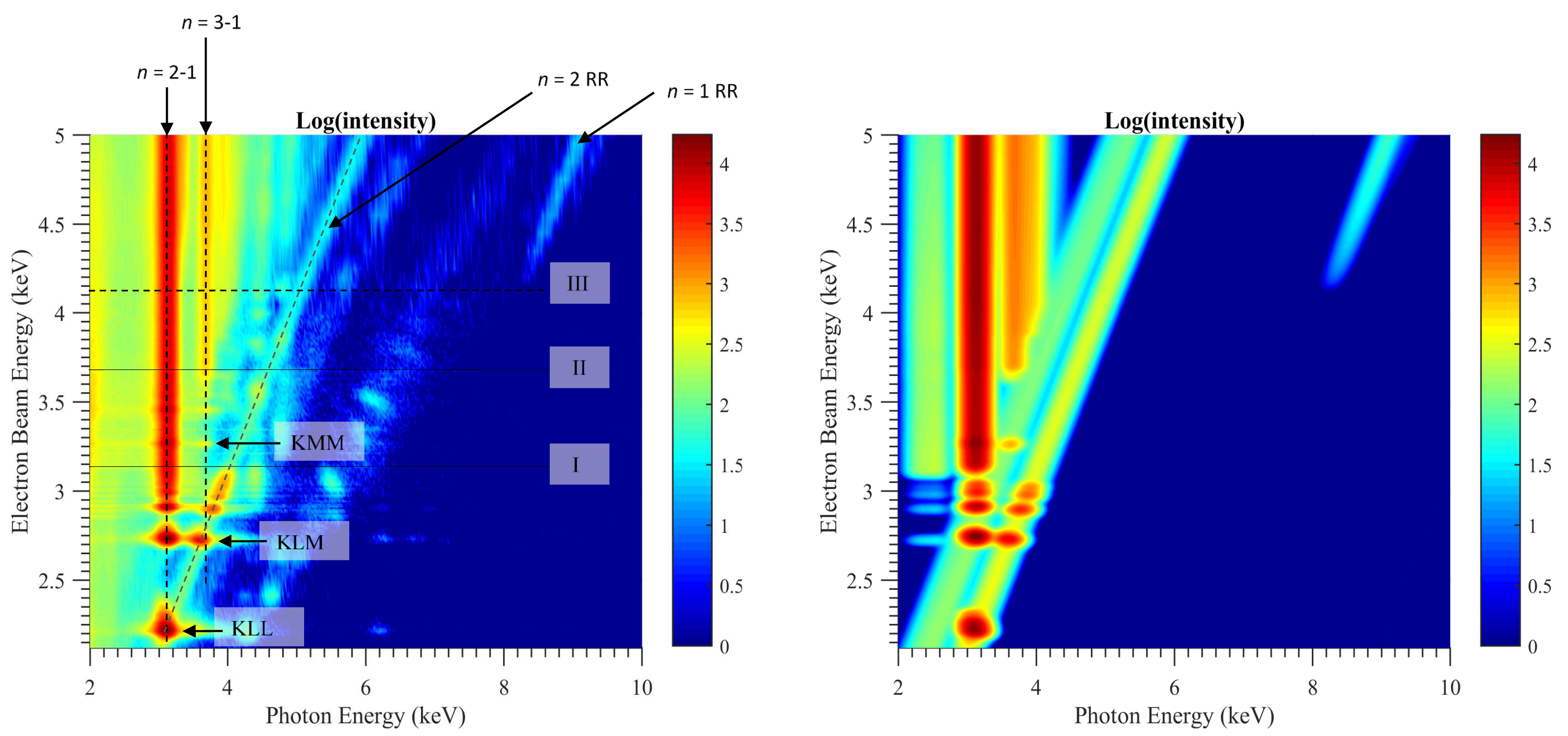}
\caption{(left) HPGe measured photon energies and intensities (counts) at electron beam energies between 2.120 keV and 5.0 keV. Dashed vertical lines highlight the signals of \emph{n}  = 2 $\rightarrow$ 1 and \emph{n}  = 3 $\rightarrow$ 1 electron transitions. Diagonal dashed line was added to denote radiative recombination into the \emph{n}  = 2 shell. Solid horizontal lines were added at the \emph{n}  = 2 $\rightarrow$ 1 (I) and \emph{n}  = 3 $\rightarrow$ 1 (II) He-like Ar direct excitation thresholds. Dashed horizontal line indicates the ground state He-like Ar ionization energy (III). (right) Corresponding data produced from the collisional-radiative model NOMAD.}
\label{fig:surf}
\end{figure}

Spectra obtained from the 3-minute measurements taken with the HPGe detector are plotted at each electron beam energy as shown in Fig. \ref{fig:surf}(left). The plot highlights some of the atomic processes occurring inside the EBIT including radiative recombination (RR), resonant DR, and direct excitation (DE). These processes present themselves in Fig. \ref{fig:surf} as diagonal lines, intense spots, and vertical lines respectively. Important Ar features have been labeled in Fig. \ref{fig:surf}(left). Weak features appearing between the \emph{n} = 2 and \emph{n} = 1 RR diagonals originate from contaminant trapped ions, including barium as previously discussed.

\begin{figure}[ht!]
\plotone{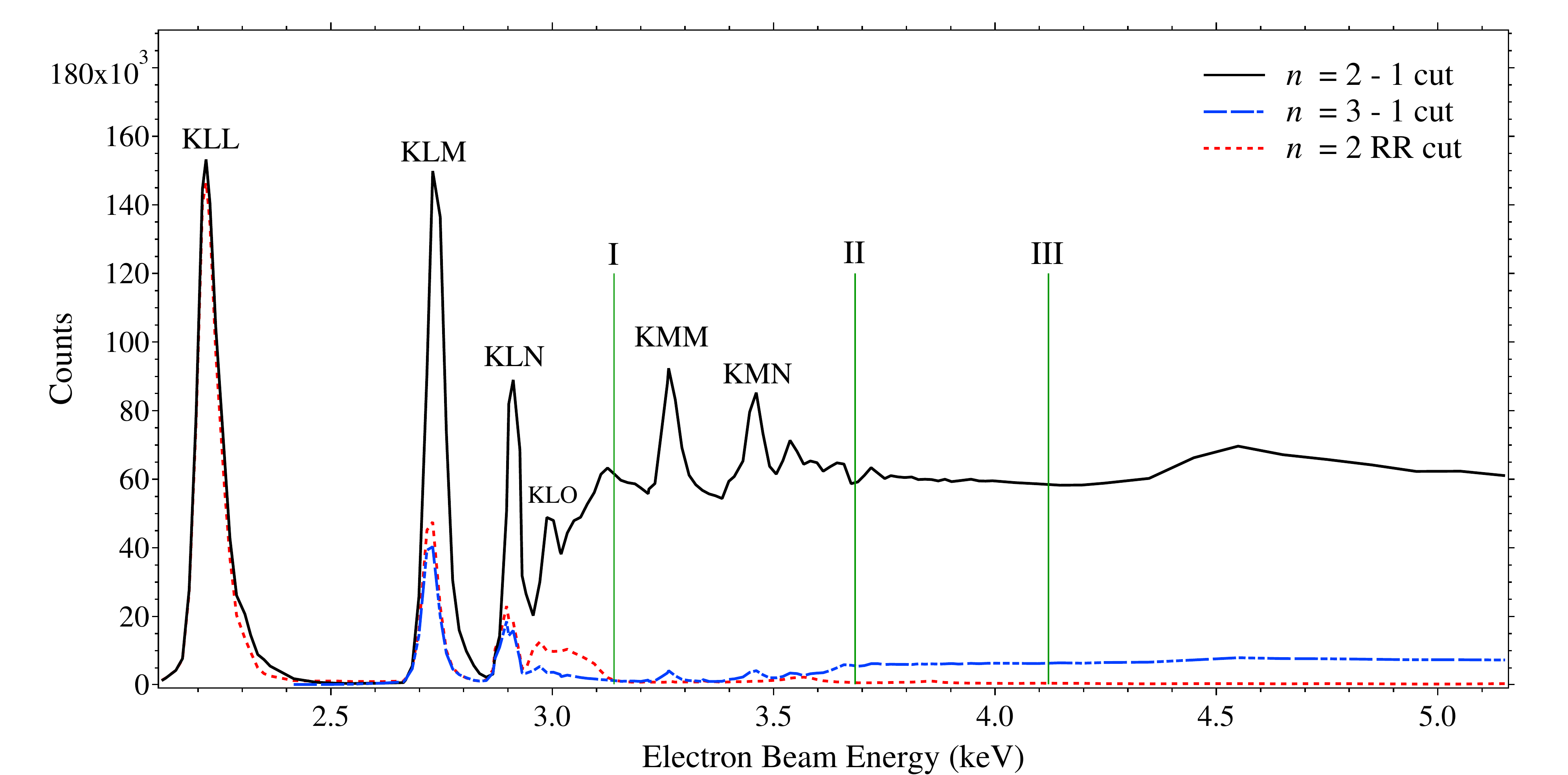}
\caption{Cuts projected onto the vertical axis from Fig. \ref{fig:surf}(left). Top thick solid curve shows \emph{n} = 2 $\rightarrow$ 1 cut. Lower blue thin curve shows \emph{n} = 3 $\rightarrow$ 1 cut. Red dotted curve shows counts from radiative recombination to the \emph{n} = 2 shell.}
\label{fig:cuts}
\end{figure}

Cuts taken through the diagonal \emph{n} = 2 RR line and down the \emph{n} = 2 $\rightarrow$ 1 and \emph{n} = 3 $\rightarrow$ 1 measured Ar lines of Fig. \ref{fig:surf}(left) were projected onto the vertical axis for a more comprehensive view as shown in Fig. \ref{fig:cuts}. For reference the \emph{n} = 2 $\rightarrow$ 1 and \emph{n} = 3 $\rightarrow$ 1 DE thresholds and the He-like ionization energy have been labeled as I, II, and III respectively in Fig. \ref{fig:surf}(left) and Fig. \ref{fig:cuts}. 

The 1s$^2$\emph{nl}$^\prime$ - 1s2\emph{lnl}$^\prime$ DR transitions in Li-like Ar are seen in the \emph{n} = 2 $\rightarrow$ 1 cut of Fig. \ref{fig:cuts} as sharp peaks below the \emph{n} = 2 $\rightarrow$ 1 DE energy threshold. Above this threshold, the He-like direct excitation is enhanced by KMM and KMN resonances. This results from L-shell Auger decay \citep{Ali91,Smith96}. The \emph{n} = 2 RR cut exposes the higher \emph{n} counterpart of the 1s$^2$2\emph{l} - 1s2\emph{lnl}$^\prime$  DRs.

\subsection{High Resolution Results} \label{subsec:highres}
\begin{figure}[ht!]
\plotone{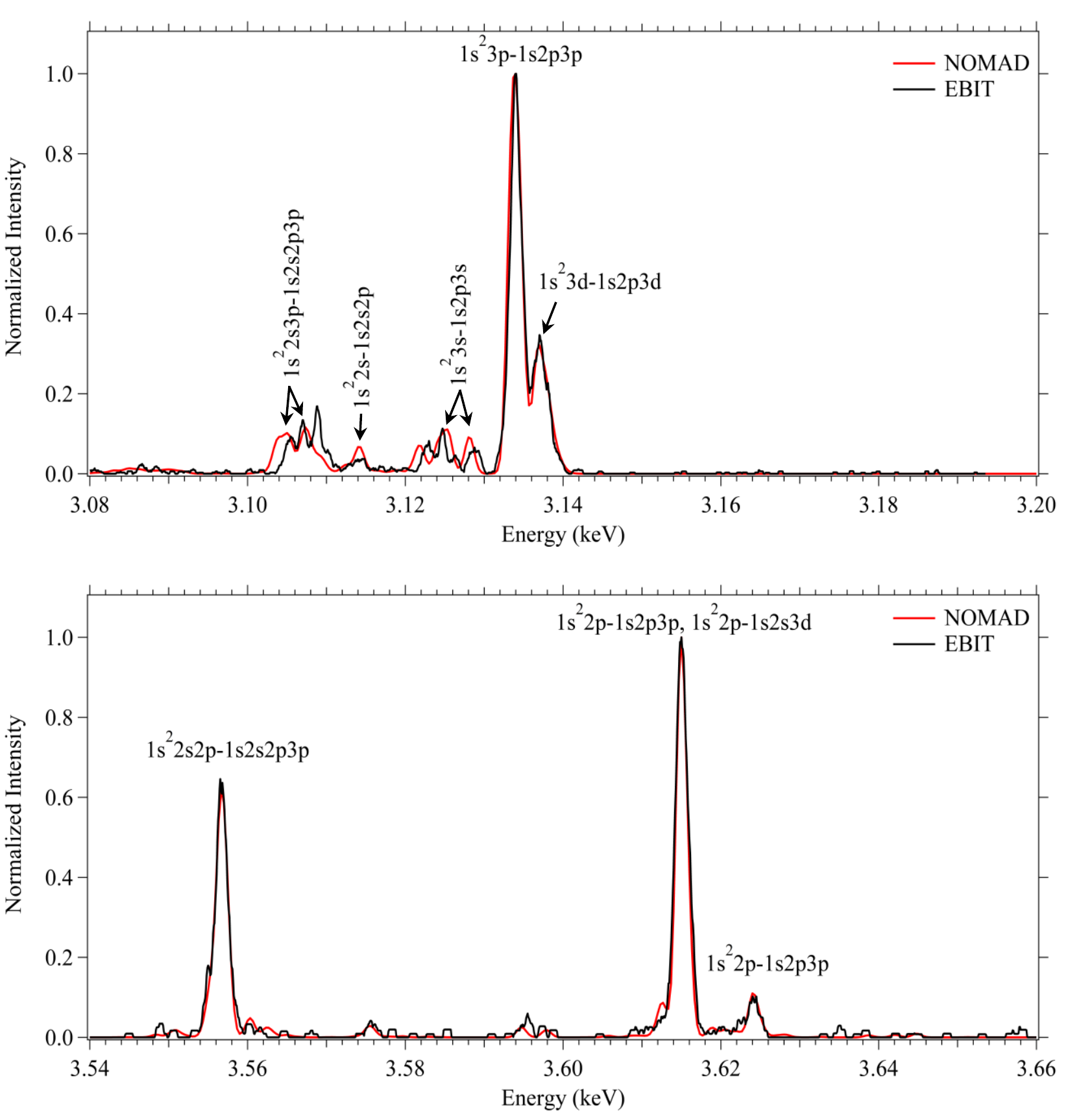}
\caption{EBIT and theoretical spectra at an electron beam energy of 2.730 keV. Photon energies between (top) 3.08 keV to 3.20 keV and (bottom) 3.54 keV to 3.66 keV, correspond to \emph{n} = 2 $\rightarrow$ 1 and \emph{n} = 3 $\rightarrow$ 1 transitions, respectively.}
\label{fig:exphr}
\end{figure}

Argon spectra measured with the high-resolution crystal spectrometer at the electron beam energy corresponding to a maximum intensity of the \emph{n} = 3 $\rightarrow$ 1 transition of the KLM resonance is shown as the solid black curve in Fig. \ref{fig:exphr}. Measurements at the KLM resonance energy were collected with a total dwell time of 15 minutes.

The detailed structure of the \emph{n} = 2 $\rightarrow$ 1 DR transitions with a spectator electron at \emph{n} = 3 in Li-like Ar is seen between 3.100 keV and 3.150 keV, while the \emph{n} = 3 $\rightarrow$ 1 transitions, with a spectator at \emph{n} = 2, are seen between 3.600 keV and 3.650 keV. The spectrum also shows corresponding lines from lower charge-states, in particular around 3.560 keV, very close in energy to the reported unidentified line as discussed in the following sections. Features have been labeled with the strongest lines for more detailed identifications.     

\subsection{Collisional-Radiative Modeling of the EBIT Plasma} \label{subsec:CREBIT}
The collisional-radiative package NOMAD \citep{Ralchenko01}, which allows for an arbitrary electron energy distribution function,  was used to calculate the ionization balance, level populations, and line intensities of the EBIT plasma. The NOMAD code uses atomic data from external sources to solve the steady-state rate equations. To this end, the flexible atomic code (FAC) \citep{Gu08} was used to calculate atomic structure, transition rates, and collisional cross-section data. Charge-exchange occurring between trapped Ar ions and neutral atoms, which can shift the charge state balance towards lower charge states, was included in the rate equations as the term: $n_0 v_0 \sigma_{CX}$, where $n_0$ is the density of neutrals in the trap, $v_0$ is the relative velocity between Ar ions and neutrals, and $\sigma_{CX}$ is the charge-exchange cross-section \citep{Ralchenko08,Ralchenko11}. Since $n_0$ and $v_0$ are not well known, the product $n_0 v_0$ was used as the only free parameter in the model. 

The simulated spectra were compared with measurements to understand the charge-state balance and correctly identify measured lines. This method has been used in previous works to accurately identify emission features from highly charged ions in x-ray and EUV spectral regions (see, e.g., \cite{Ralchenko06, Ralchenko07,Ralchenko11,Podpaly14,Kilbane14,Reader14,Silwal17}). Many of the earlier works also provide a thorough explanation of the calculations which are omitted here.

Fig. \ref{fig:surf}(right) shows the modeled EBIT plasma convolved with a Gaussian of FWHM of 120 eV. Measured features including the intense DR resonances, direct-excitation lines, and RR diagonals are clearly reproduced, verifying our model at each electron beam energy. The theoretical spectrum, calculated at an electron beam energy of 2.730 keV and convolved with a Gaussian of FWHM of 1.4 eV, is shown with our EBIT spectra in Fig. \ref{fig:exphr}. Measured KLM DR features seen in our EBIT spectra are well reproduced, providing additional confidence in our model. 

It is important to note that the emission produced by the uni-directional electron beam in the EBIT can be polarized and anisotropic. Furthermore, the crystal spectrometer is sensitive to the polarization (see, e.g., \cite{Henderson90,Beiersdorfer96,Takacs96}). The agreement seen between our modeled and experimental spectra, particularly at the 1s$^2$2p - 1s2p3p and 1s$^2$2s2p - 1s2s2p3p DR peaks of interest, suggest that polarization effects from these sources were not significant and were not considered for the DR analysis in this work. Additional efforts are currently underway to investigate polarization of DR transitions in Li-like Ar.

\subsection{Spectra from Collisional-Radiative Maxwellian Models} \label{subsec:CRMax}
\begin{figure}[ht!]
\plotone{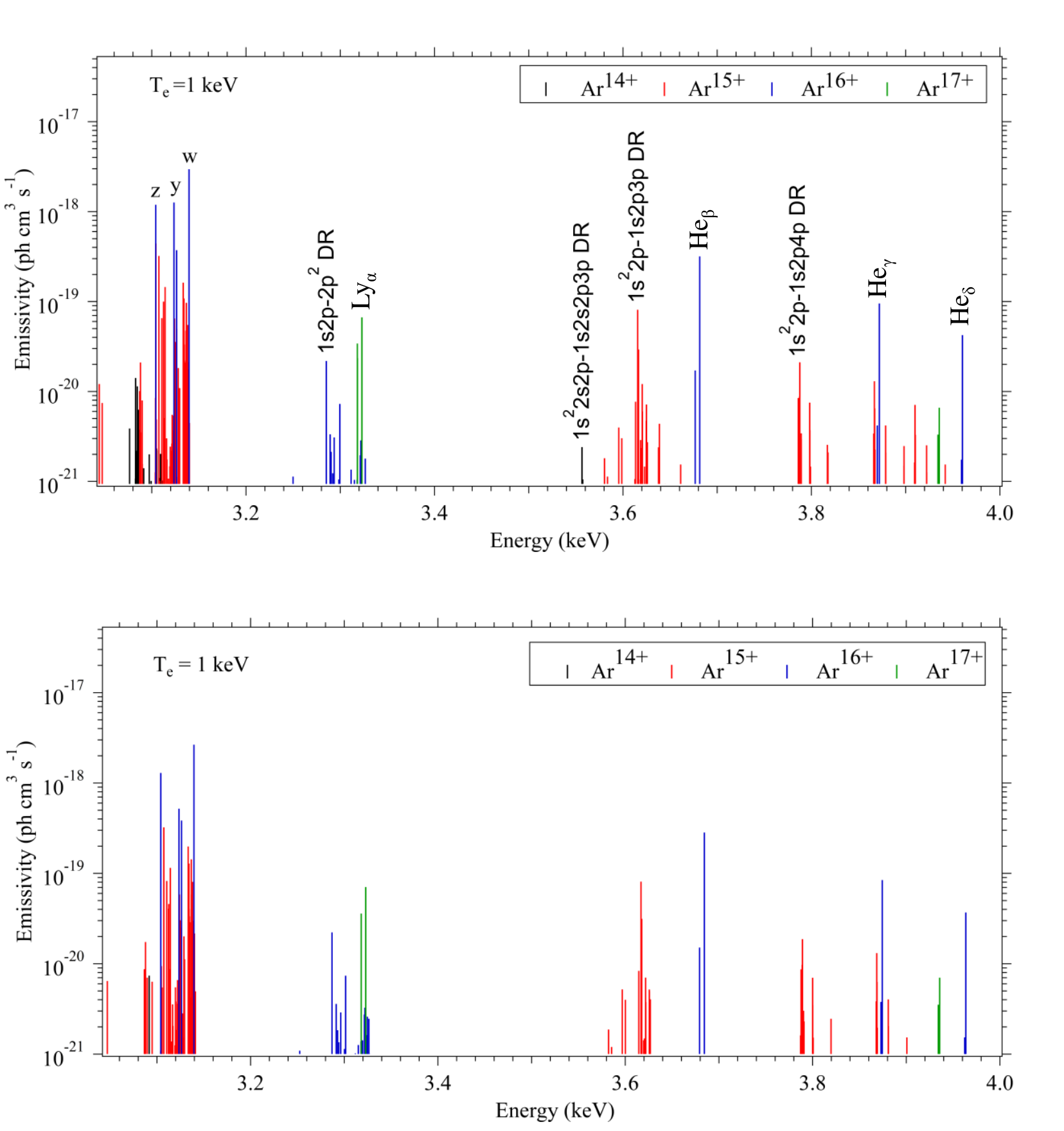}
\caption{Ar spectrum calculated at T$_e$ = 1 keV with (top) NOMAD and (bottom) APEC.}
\label{fig:onekev}
\end{figure}

The ions present in the EBIT trap are produced and excited by a quasi-monoenergetic electron beam, producing a non-Maxwellian plasma; however, the hot intracluster medium of galaxy clusters, responsible for producing the majority of the emission, is assumed to follow a Maxwellian distribution. To predict the importance of experimentally observed features under these conditions, we applied our CR model, which accurately reproduces measured spectra, to a Maxwellian distributed electron energy distribution with electron temperature T$_e$. The calculated Ar spectra at T$_e$ = 1 keV detailed in Fig. \ref{fig:onekev}(top) includes strong He-like direct excitation features, Li-like DR transitions, and weaker Be-like DR transitions. The two strong Li-like DR transitions of interest mentioned in \cite{Bulbul14} are observed near 3.62 keV along with a number of weaker Li-like DR transitions. Close in energy to the unidentified line, near 3.57 keV, we see lower charge-state Be-like Ar DR transitions and additional Li-like DR transitions.

AtomDB is an atomic database that includes the Astrophysical Plasma Emission Database (APED) containing fundamental atomic data such as wavelengths, radiative transition rates, and electron collisional excitation rate coefficients. AtomDB also includes the spectral models output from the Astrophysical Plasma Emission Code (APEC) \citep{Smith01}. APEC uses the data from APED to calculate line emissivities for optically-thin plasmas in collisional ionization equilibrium. In Fig. \ref{fig:onekev}(bottom) we utilized APEC to calculate the same Ar spectrum at T$_e$ = 1 keV for comparison with NOMAD. Lines with an emissivity below 10$^{-20}$ (ph cm$^3$s$^{-1}$) are typically not included as individual emission features in the AtomDB data but are instead included in a pseudo-continuum consisting of weak lines. For our calculation, the emissivity cutoff was lowered to 10$^{-22}$ ph cm$^3$s$^{-1}$, and as a result, the calculated spectra from AtomDB is seen to have more weak lines when compared to our calculated spectra in Fig. \ref{fig:onekev}(top). However, the strongest lines show the same overall structure.

\begin{figure}[ht!]
\plotone{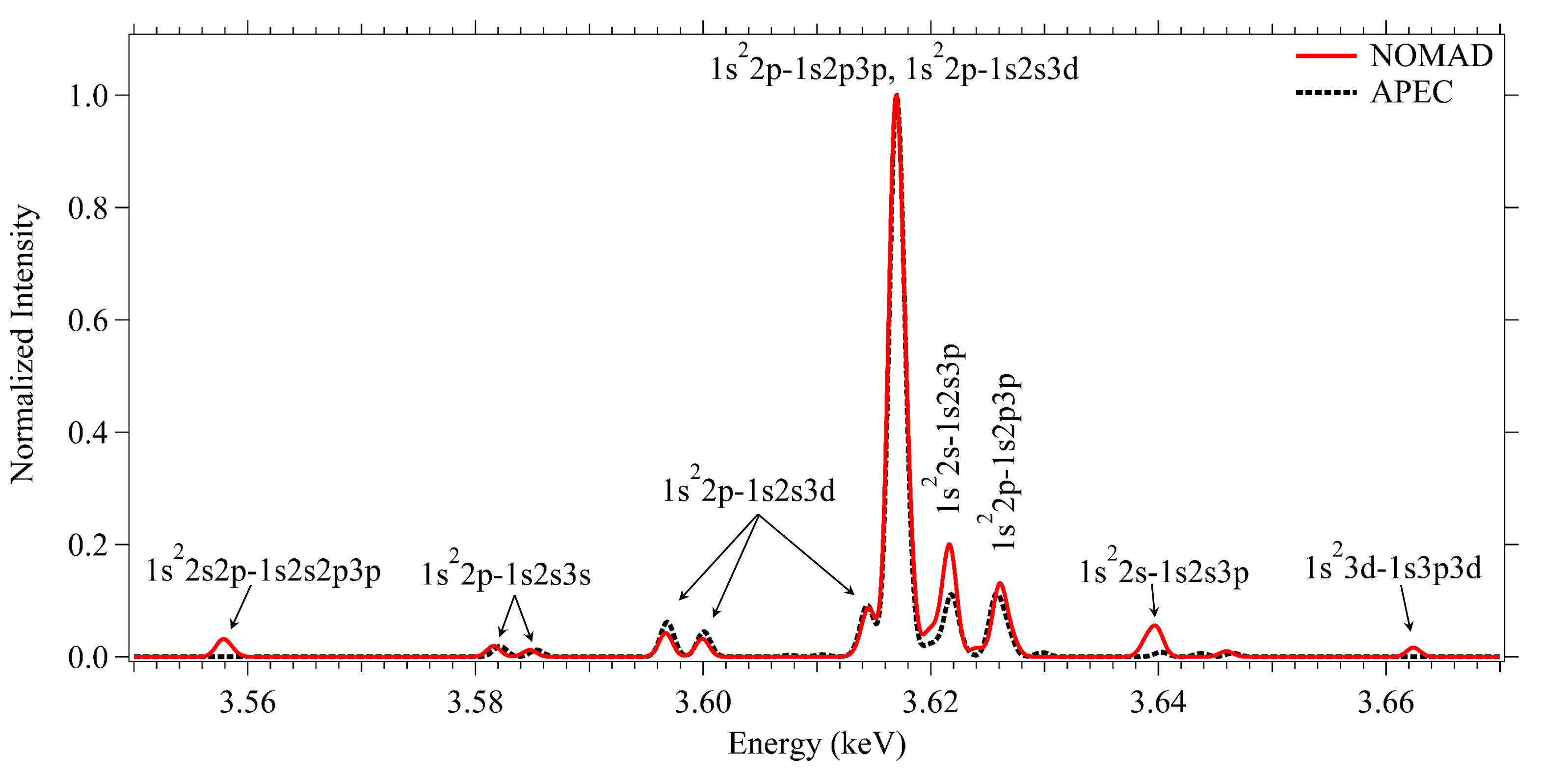}
\caption{Comparison of the DR spectra produced with APEC and spectra produced with NOMAD near the unidentified line at 3.57 keV.}
\label{fig:twoeVconv}
\end{figure}

Focusing only on DR transitions, we overlaid our NOMAD DR spectra, convolved with a 1.4 eV FWHM Gaussian to match the resolution of the crystal spectrometer, with that produced by APEC in the energy region of interest. The spectra were normalized to the strongest DR feature near 3.616 keV. The DR intensities in Fig. \ref{fig:twoeVconv} are generally in good agreement with a few features missing in the APEC spectra near 3.56 keV, 3.62 keV, 3.64 keV, and 3.66 keV. While much of the missing emission is due to a forest of weak Be-like lines missing from the database AtomDB, we also found a few missing or underestimated intensities from Li-like transitions also contributing. In particular the 1s$^2$2s - 1s2s3p transitions at 3.62 keV and 3.64 keV are much stronger in our model and are partially responsible for the missing emission. The missing emission near 3.56 keV is solely due to Be-like transitions (discussed further in section \ref{sec:discussion}), and the 3.66 keV line originates from missing 1s$^2$3d - 1s3p3d DR transitions.

\section{DISCUSSION} \label{sec:discussion}
The astrophysical atomic database AtomDB was used in the analysis of the stacked spectra of galaxy clusters \citep{Bulbul14}. Strong Ar emission lines were fit along with a few weaker features, including the He-like Ar DR satellites listed in AtomDB at energies of 3.618 keV and 3.617 keV and relative intensities of 0.39 and 1, respectively. Though these two lines are not fully resolved, even in our measured high-resolution spectra, we do see the blended line in Fig. \ref{fig:exphr}. The projected cuts taken through the EBIT data plot shown in Fig. \ref{fig:cuts} highlight strong DR resonances and show the relative strength of the \emph{n} = 2 $\rightarrow$ 1 to the \emph{n} = 3 $\rightarrow$ 1 satellite transitions of interest at the KLM DR peak.

In their report, \cite{Bulbul14} calculated the maximum emissivity of the 3.62 keV Ar DR line to be 4 $\%$ of the He-like Ar triplet at 3.12 keV, at T$_e$ = 0.7 keV. They also note that if the unidentified galaxy cluster line results from the 3.62 keV DR resonance, then the flux would need to be increased by a factor of 30 from the current AtomDB estimate. As a check, we looked at the 3.62 keV DR resonance feature in the NOMAD and APEC spectra and compared its emissivity to the He-like triplet. In agreement with \cite{Bulbul14}, at T$_e$ = 1 keV the 3.62 keV DR is roughly 2 $\%$ of the Ar triplet in both spectra. Given that the relative intensity ratio of the strong He-like lines to the 3.62 keV DR line is also comparable between our EBIT measurements and calculated spectra, we conclude that the He-$\beta$ DR data used in \cite{Bulbul14} are not off as much as the factor of 30 needed for known atomic physics to resolve the problem.

During our investigation, we measured an interesting feature very close in energy to the unidentified line near 3.56 keV. This line, seen in the measured EBIT spectrum and replicated in our NOMAD calculated spectrum (Fig. \ref{fig:exphr}), has an intensity comparable to the Ar DR satellite feature near 3.62 keV.  Using the NOMAD model, we were able to identify this as 1s$^2$2s2p - 1s2s2p3p electric dipole DR transitions from Ar$^{14+}$ with an approximate energy of 3.557 keV. AtomDB does not include DR satellite lines for Ar$^{15+}$ recombining to Ar$^{14+}$, therefore the 3.557 keV feature was not included in \cite{Bulbul14}, and it is not in the AtomDB spectra in Fig. \ref{fig:onekev}(bottom) and Fig. \ref{fig:twoeVconv}. It can clearly be seen in our T$_e$  = 1 keV NOMAD spectrum (Fig. \ref{fig:onekev}(top) and Fig. \ref{fig:twoeVconv}). 

\begin{figure}[ht!]
\plotone{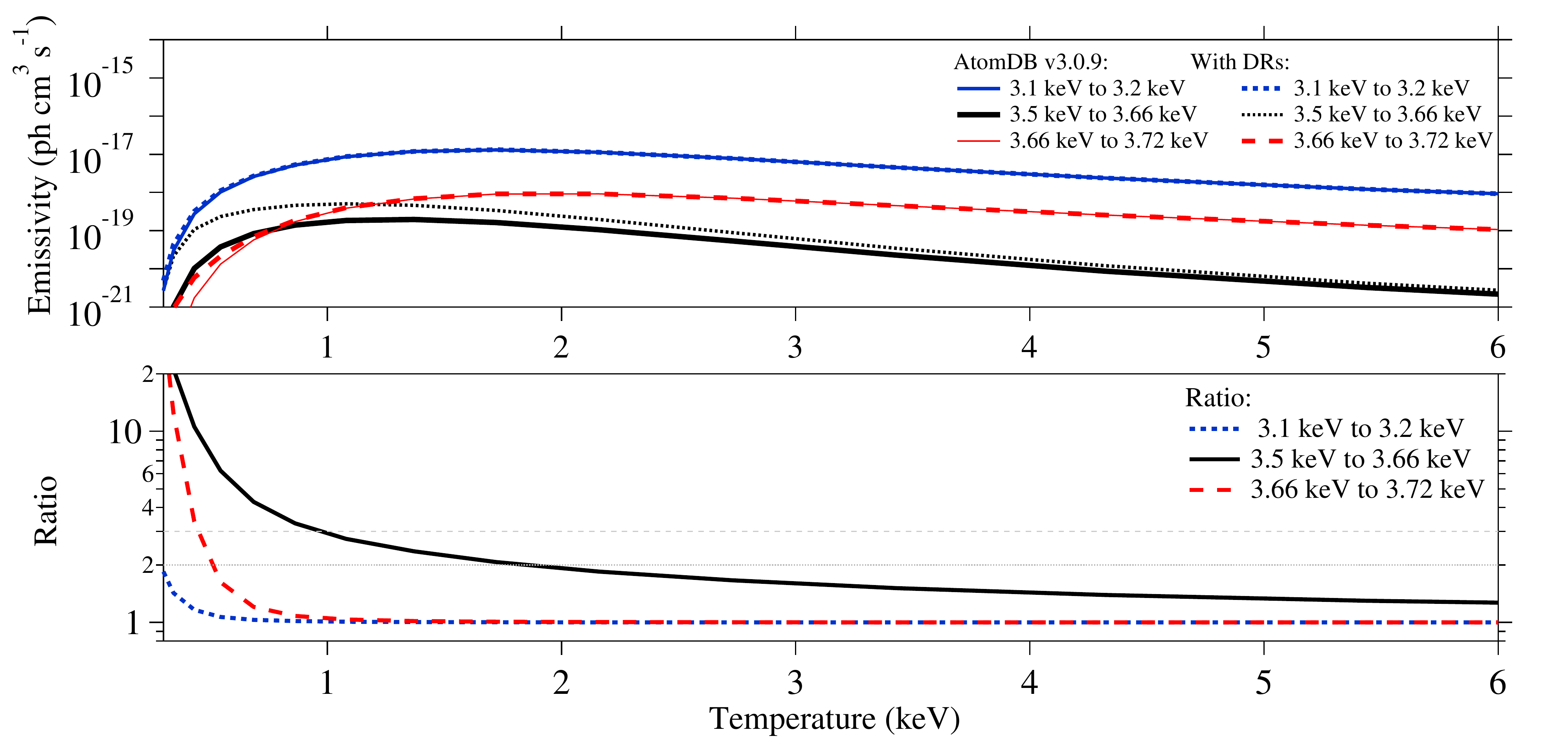}
\caption{(top) Total emissivity produced by AtomDB v.3.0.9 in the energy bands: 3.1 keV to 3.2 keV, 3.5 keV to 3.66 keV, and 3.66 keV to 3.72 keV at temperatures below 6 keV are shown as solid lines. Total emissivity produced by AtomDB, with Ar$^{15+}$ DR features included, are shown as dotted lines. (bottom) Ratio of the total flux with new DR lines included to the original flux (not including Ar$^{15+}$ DR lines) are shown for each energy band.}
\label{fig:ratios}
\end{figure}

Using FAC we produced data for 1s2\emph{l}2\emph{l}$^\prime$2\emph{l}$^{\prime\prime}$ and 1s2\emph{l}2\emph{l}$^\prime$3\emph{l}$^\prime$ DR satellite lines between 3.075 keV and 3.672 keV. This data was added to AtomDB and the ratio of the flux with and without the new lines was evaluated in three energy bands and over a range of electron temperatures. The three energy bands include: 3.1 keV to 3.2 keV (corresponding to the Ar$^{16+}$ He-$\alpha$ complex), 3.66 keV to 3.72 keV (at Ar$^{16+}$ He-$\beta$), and 3.5 keV to 3.66 keV (where the unidentified line, Ar$^{16+}$, and Ar$^{15+}$ DR lines lie). As shown in Fig. \ref{fig:ratios}, the added data has minimal effects on the He-$\alpha$ and He-$\beta$ complexes, as these lines were already very bright. However, the new DR data leads to significant enhancement of the DR feature around 3.6 keV, especially at lower temperatures. 

\begin{figure}[ht!]
\plotone{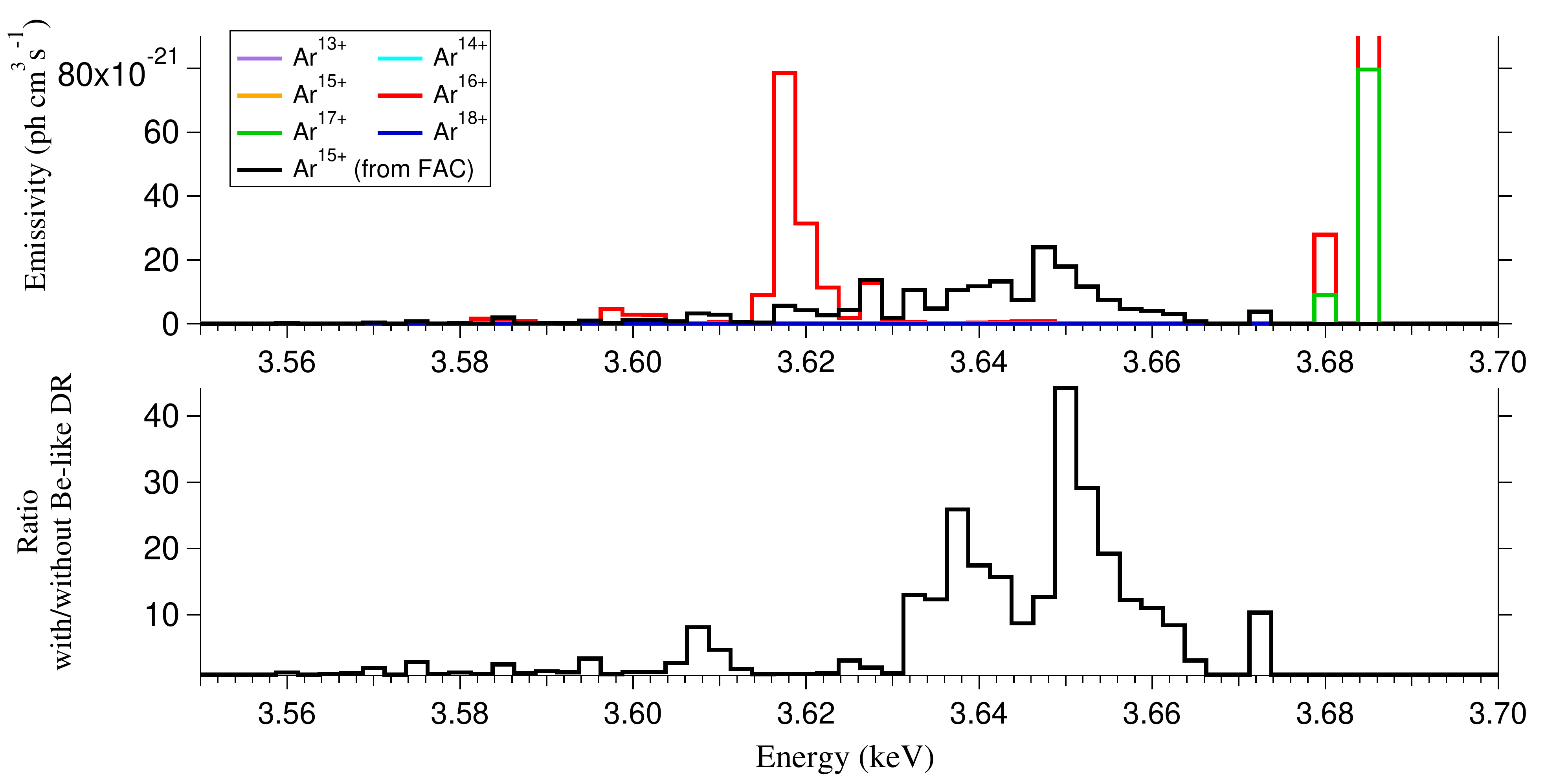}
\caption{(top) Spectra of Ar ions at T$_e$ = 1.72 keV. Ar$^{15+}$ DR lines added from FAC shown in black. (bottom) Ratio of the total flux in each energy bin with the new DR lines included to the original flux (not including Ar$^{15+}$ DR lines).}
\label{fig:onepsevenkeV}
\end{figure}

In Fig. \ref{fig:onepsevenkeV}(top) the Ar emissivity was calculated at T$_e$ = 1.72 keV using the original AtomDB data and again with the newly included DR data. The emission is broken up for each Ar charge state, demonstrating the lack of Ar$^{15+}$ features in the original spectrum. The new Ar$^{15+}$ DR features are seen predominately around 3.64 keV. Their effect is further highlighted in Fig. \ref{fig:onepsevenkeV}(bottom) where the ratio of the total emissivity with and without the features is calculated for each energy bin. This produces a maximum factor of 44 increase in emissivity around 3.65 keV.  

\cite{Bulbul14} find a range of temperatures for different components used to model their plasma. The lowest of these is T$_e$ = 2.0 keV for the ``Excluding Nearby Clusters'' sample. At this temperature, the new DR data enhances the flux in the 3.5 keV to 3.66 keV band by a factor of 2. While significant, this is much smaller than the factor of 30 which \cite{Bulbul14} state is required for this line to explain the 3.55 keV feature.

In a recently released preprint, \cite{Bulbul18} (BUL18) report EBIT experiments in a similar vein to these, aiming to measure the effect of the Ar DR emission. Their results are similar to ours in that they also find that their measured Ar DR is more intense than allowed for in AtomDB and therefore in \cite{Bulbul14}, but not by the factor of 30 required for Ar DR to explain the unidentified feature. In particular at T$_e$ = 1.74 keV, BUL18 report a factor of 2.6 in missing flux in the 3.54 keV to 3.645 keV range (from BUL18, Table 4) when comparing EBIT measurements to spectra produced with AtomDB v3.0.8. They added DR data from \cite{Beiersdorfer95} Tables V and VI including 1s3\emph{l}3\emph{l}$^\prime$ data spanning 3.645 keV to 3.680 keV and 1s2s2\emph{l}3\emph{l}$^\prime$ data between 3.145 keV and 3.588 keV. With the added lines a better fit at 3.55 keV to 3.59 keV was obtained, but no improvement between 3.6 keV and 3.65 keV was observed. BUL18 mention that the additional lines cannot explain the extra flux they measured in the Ar$^{16+}$ He-$\beta$ DR lines. Since the Ar$^{15+}$ DR lines added in their work only spanned 3.145 keV to 3.588 keV, conclusions cannot be made regarding their effect on the total flux in this region.

As previously discussed in this work, we added Be-like DR data to AtomDB covering a wider energy range (3.075 keV to 3.672 keV). At T$_e$ = 1.72 keV (close to the BUL18 temperature) the addition of these lines produced a factor of 2 increase in the total flux between 3.5 keV and 3.66 keV. As demonstrated in Fig. \ref{fig:onepsevenkeV}, we saw the largest increase in flux between 3.63 keV and 3.67 keV suggesting these lines account for a large portion of missing flux reported by BUL18 in Table 4. The largest discrepancy they report between experiment and AtomDB is a factor of 10.7 difference in flux in the energy region between 3.630 keV and 3.645 keV. Inclusion of the Be-like data in this work increases the flux in this energy region by a factor of 14.5 at T$_e$ = 1.72 keV. Finally, as discussed in section \ref{subsec:CRMax}, we also found 1s$^2$2s - 1s2s3p transitions at 3.62 keV and 3.64 keV that were either missing or greatly underestimated in AtomDB. Amending these issues will add more to the missing flux in this region.

\section{CONCLUSIONS} \label{sec:conclusions}
Searching for possible atomic origins of the unidentified line in the stacked spectra of galaxy clusters \citep{Bulbul14}, we measured x-ray emission from Ar ions at the NIST EBIT facility. The excellent agreement shown between our EBIT and NOMAD modeled non-Maxwellian spectra provides a high level of confidence in the atomic data used in our model. In comparing a T$_e$ = 1 keV Maxwellian-distributed spectra produced by the NOMAD code to that produced with AtomDB/APEC we find good agreement in the line intensity ratio of the He-like triplet to the 3.62 keV DR, confirming that the Ar$^{16+}$ DR is not off by the factor of 30 required to explain the unidentified feature. We also found that the AtomDB spectra has significant emission missing in the energy region near the unidentified line due to Ar$^{15+}$ DR features. Including missing Ar$^{15+}$ DR data in AtomDB resulted in a factor of 2 increase in the flux between 3.5 keV and 3.66 keV at T$_e$ = 1.72 keV. There are also a number of Ar$^{16+}$ DR transitions missing or underestimated in the AtomDB data that show up near 3.64 keV and 3.62 keV in Fig. \ref{fig:twoeVconv}. Combined, these features contribute to a significant amount of emission which was not accounted for in AtomDB and therefore not in the \cite{Bulbul14} work. These missing or inaccurate DR lines also account for the missing emission reported in BUL18 in this energy region. 

Finally, while charge-states lower than He-like Ar may not contribute significantly to individual galaxy cluster emission (which are typically at much higher temperatures where these low charge-states are less abundant), they may be important in lower temperature astrophysical objects, in non-Maxwellian plasma sources, and in stacked spectra where weak features can be greatly enhanced. This was clearly demonstrated by comparing spectra from our controlled EBIT plasma to modeled spectra. However, in astrophysical plasmas containing multiple elements, charge states, and electron energies the results may be more subtle and lead to physical misinterpretations making the inclusion and accuracy of this data important.  

\section{Acknowledgments} \label{sec:ack}
Authors would like to thank Dipti, Nancy Brickhouse, and Randall Smith for their detailed feedback on the draft. This work was partially funded by the NIST Grant Award Numbers 70NANB16H204 and 70NANB18H284 of the Measurement Science and Engineering (MSE) Research Grant Programs.


\software{NOMAD \citep{Ralchenko01}, FAC \citep{Gu08}, AtomDB \citep{Foster12}, APEC \citep{Smith01}}

\end{document}